\documentclass[12pt]{article}
\usepackage{amssymb}
\usepackage{graphicx}
\usepackage{epstopdf}
\usepackage{amsmath}
\usepackage{color}
\usepackage{lscape}
\usepackage{booktabs}

\allowdisplaybreaks

\begin{document}

\baselineskip=20pt

\renewcommand{\theequation}{\arabic{section}.\arabic{equation}}

\newcommand{\Title}[1]{{\baselineskip=26pt	\begin{center} \Large \bf #1 \\ \ \\ \end{center}}}

\newcommand{\Author}{\begin{center}
	\large \bf
	Pei Sun${}^{a}$, Jintao Yang${}^{b,c}$, Yi Qiao${}^{b}$\footnote{Corresponding author: qiaoyi\_joy@foxmail.com}, Junpeng Cao${}^{b,c,d,e}$\footnote{Corresponding author: junpengcao@iphy.ac.cn} and Wen-Li Yang${}^{a,e,f,g}$\footnote{Corresponding author: wlyang@nwu.edu.cn}
\end{center}}

\newcommand{\Address}{\begin{center}
    ${}^a$ Institute of Modern Physics, Northwest University, Xian 710127, China\\
	${}^b$ Beijing National Laboratory for Condensed Matter Physics, Institute of Physics, Chinese Academy of Sciences, Beijing 100190, China\\
    ${}^c$ School of Physical Sciences, University of Chinese Academy of Sciences, Beijing 100049, China\\
	${}^d$ Songshan Lake Materials Laboratory, Dongguan, Guangdong 523808, China\\
	${}^e$ Peng Huanwu Center for Fundamental Theory, Xian 710127, China\\
	${}^f$ School of Physics, Northwest University, Xian 710127, China\\
	${}^g$ Shaanxi Key Laboratory for Theoretical Physics Frontiers, Xian 710127, China
\end{center}}

\Title{Scattering matrix of elementary excitations in the antiperiodic $XXZ$ spin chain with $\eta=\frac{i\pi}{3}$}
\Author

\Address

\vspace{1truecm}

\begin{abstract}

We study the thermodynamic limit of the antiperiodic XXZ spin chain with the anisotropic parameter $\eta=\frac{\pi i}{3}$.
We parameterize eigenvalues of the transfer matrix by their zero points instead of  Bethe roots. We obtain  patterns of the distribution of zero points. Based on them, we calculate the ground state energy and the elementary excitations in the thermodynamic limit.
We also obtain the two-body scattering matrix of elementary excitations. Two types of elementary excitations and three types of scattering processes are discussed in detailed.

\vspace{1truecm}

\end{abstract}

\newpage
\section{Introduction}
\label{introduction}

The $XXZ$ spin chain with antiperiodic boundary condition is a typical quantum integrable model with $U(1)$-symmetry broken \cite{1,2,3,4,5,6}, which has many applications in the non-equilibrium statistical
physics \cite{Q3-6,Q3-7}, quantum magnetism \cite{Q3-5} and high energy physics \cite{Q3-1}.
The integrability of the model is guaranteed by the trigonometric solution, the six-vertex $R$-matrix, of the Yang-Baxter equation \cite{8,9,10,11}.
Due to the broken of $U(1)$ symmetry, the properties of the system is very different from that of the periodic XXZ spin chain.
Furthermore, the coordinate and algebraic Bethe ansatz could not be applied \cite{12,12-1,13,14,15}.
Recently, with the help of the new proposed off-diagonal Bethe ansatz, the exact solution including the energy spectrum and helical eigenstates of the system are obtained \cite{16,19,20,21}.
The eigenvalue of the transfer matrix is expressed by the inhomogeneous $T-Q$ relations, where the Bethe root should satisfy the inhomogeneous Bethe ansatz equations (BAEs).
As a result, the thermodynamic limit is hard to be reached because that tradition thermodynamic Bethe ansatz does not work.

In order to solve this problem, the $t-W$ scheme is proposed \cite{Qiao20,Qiao21}. The main idea is that the eigenvalue of the transfer matrix can also be characterized by its zero points.
It is found that the zero points satisfy the homogeneous BAEs. Then we can define the densities of state and study the exact results in the thermodynamic limit.
The next task is that we should derive the exact thermodynamic quantities of the system at finite temperature, where the patterns of zero points distribution
should be addressed first.

In this paper, we study the zero points BAEs of the antiperiodic $XXZ$ spin chain with the anisotropic parameter $\eta=\frac{i\pi}{3}$.
We obtain the patterns of zero points distribution of the eigenvalue of the transfer matrix.
Based on them, we calculate the physical quantities such as the ground state energy and elementary excitations in the thermodynamic limit.
We also obtain the two-body scattering matrix of elementary excitations. The three kinds of scattering processes are discussed exactly.

The reason that we choose the specific point of $\eta=\frac{i\pi}{3}$ is as follows. As we mentioned above, the eigenstates of the antiperiodic $XXZ$ spin chain are helical \cite{20,21}.
At present, people do not find the explicit form of good quantum numbers
characterizing the eigenstates of the quantum integrable models without $U(1)$ symmetry. However, if the anisotropic parameter $\eta=\frac{i\pi}{3}$,
roots of the transfer matrix satisfy a homogeneous BAEs (see (\ref{BAE-2}) below), which allows us to use the conventional Bethe ansatz to obtain the root patterns.
Thus we can write out the quantum numbers explicitly. Based on them,
we can calculate the physical quantities of the system in the thermodynamic limit.

This paper is organized as follows. In section \ref{sec2}, we briefly review the antiperiodic $XXZ$ spin chain and show its integrability. In section \ref{Interaction-case}, we derive
the exact solution of the system with the anisotropic parameter $\eta=\frac{i\pi}{3}$. The eigenvalue of transfer matrix is parameterized by its zero points. The analytical results are checked by the numerical calculation.
In section \ref{the gs}, the pattern of zero points and the quantum numbers at the ground state are obtained. Based on them, we obtain the ground state energy in the thermodynamic limit.
In section \ref{the ee}, by analyzing the distribution of zero points in the complex plane, we obtain two kinds of elementary excitations. The corresponding excited energies are calculated.
In section \ref{scattering matrix}, we derive the two-body scattering matrix of elementary excitations. The three types of scattering processes are discussed.
Concluding remarks are given in section \ref{sec:concluding remarks}.

\section{Integrability of the model}\label{sec2}
\setcounter{equation}{0}

The antiperiodic XXZ spin chain is characterized by the Hamiltonian
\begin{eqnarray}\label{Ham}
H=-\sum_{j=1}^N\big[ \sigma_j^x \sigma_{j+1}^x + \sigma_j^y \sigma_{j+1}^y +\cosh\eta \sigma_j^z \sigma_{j+1}^z \big],
\end{eqnarray}
where $N$ is the number of sites, $\sigma_j^\alpha $ is the Pauli matrix at $j$-th site along the $\alpha$-direction and $\eta$ is the anisotropic parameter.
The boundary condition is the twisted one, i.e.,
\begin{eqnarray}
\sigma_{N+1}^\alpha=\sigma_1^x\sigma_1^\alpha\sigma_1^x,\quad {\rm for}\quad \alpha=x, y, z.\label{Anti-periodic}
\end{eqnarray}
It is remarked  that the model possesses a $Z_2$ invariance $[H,
U]=0$ with $U=\prod_{j=1}^N\sigma^x_j$ and $U^2=1$.

The integrability of the model (\ref{Ham}) is associated with the six-vertex $R$-matrix \cite{30,31,32,33,34,35}
\begin{eqnarray}\label{R-matrix}
R_{0,j}(u)= \frac{1}{2} \left[ \frac{\sinh(u+\eta)}{\sinh \eta} (1+\sigma^z_j \sigma^z_0) +\frac{\sinh u}{\sinh \eta} (1- \sigma^z_j \sigma^z_0)   \right]+ \frac{1}{2} (\sigma^x_j \sigma^x_0 +\sigma^y_j \sigma^y_0) ,
\end{eqnarray}
where $u$ is the spectral parameter. Throughout this paper, we adopt the standard notations. For
any matrix $A\in {\rm End}(\mathbb{C})$, $A_j$ is an embedding operator
in the tensor space $\mathbb{C}^2\otimes \mathbb{C}^2\otimes\cdots$, which acts
as $A$ on the $j$-th space and as identity on the other factor
spaces. $R_{i,j}(u)$ is an embedding operator of $R$-matrix in the
tensor space, which acts as identity on the factor spaces except
for the $i$-th and $j$-th ones. Thus the $R$-matrix (\ref{R-matrix})
is defined in the auxiliary space $\mathbb{C}_0$ and the quantum space $\mathbb{C}_j$.

The $R$-matrix (\ref{R-matrix}) satisfies the Yang-Baxter equation
\begin{eqnarray}
R_{1,2}(u_1-u_2)R_{1,3}(u_1-u_3)R_{2,3}(u_2-u_3)
=R_{2,3}(u_2-u_3)R_{1,3}(u_1-u_3)R_{1,2}(u_1-u_2).\label{QYB}
\end{eqnarray}
Besides, the $R$-matrix (\ref{R-matrix}) has the following  properties
\begin{eqnarray}
&&\hspace{-1.5cm}\mbox{ Initial
condition}:\,R_{1,2}(0)= P_{1,2},\label{Int-R}\nonumber \\
&&\hspace{-1.5cm}\mbox{ Unitarity
relation}:\,R_{1,2}(u)R_{2,1}(-u)= -\xi(u)\times\,{\rm id},
\;\; \xi(u)=\frac{\sinh(u-\eta)\sinh(u+\eta)}{\sinh\eta\,\sinh\eta},\label{Unitarity}\nonumber \\
&&\hspace{-1.5cm}\mbox{ Crossing
relation}:\,R_{1,2}(u)=V_1R_{1,2}^{t_2}(-u-\eta)V_1,\quad
V_1=-i\sigma_1^y,
\label{crosing-unitarity}\nonumber \\
&&\hspace{-1.5cm}\mbox{ PT-symmetry}:\,R_{1,2}(u)=R_{2,1}(u)=R^{t_1\,t_2}_{1,2}(u),\label{PT}\nonumber \\
&&\hspace{-1.4cm}\mbox{$Z_2$-symmetry}: \;\;
\sigma^\alpha_1\sigma^\alpha_2R_{1,2}(u)=R_{1,2}(u)\sigma^\alpha_1\sigma^\alpha_2,\quad
\mbox{for}\,\,
\alpha=x,y,z,\label{Z2-sym}\nonumber \\
&&\hspace{-1.5cm}\mbox{ Quasi-periodicity}:\, R_{1,2}(u+i\pi)=-\sigma^z_1\,R_{1,2}(u)\,\sigma^z_1.\label{quasi-}
\end{eqnarray}
Here $R_{2,1}(u)=P_{1,2}R_{1,2}(u)P_{1,2}$ with $P_{1,2}$ being
the permutation operator and $t_j$ denotes transposition in
the $j$-th space.

With the help of $R$-matrix (\ref{R-matrix}), the transfer matrix of the system (\ref{Ham}) is constructed as
\begin{equation}\label{transfer_matrix}
t(u)=tr_0 \{ \sigma_0^x R_{0,N}(u-\theta_N)\ldots R_{0,1}(u-\theta_1) \},
\end{equation}
where $tr_0$ denotes partial trace over the
auxiliary space, and $\{\theta_j|j=1,\cdots,N\}$ are the generic free complex
parameters which are usually called as inhomogeneous parameters.
From the Yang-Baxter equation (\ref{QYB}), one can prove that the transfer matrices with different spectral parameters commute with each other,
\begin{equation}
[t(u),t(v)]=0.\label{ai}
\end{equation}
Thus the transfer matrix \eqref{transfer_matrix} serves as the generating functional of the conserved quantities, which ensures the integrability of the system (\ref{Ham}).
Expanding the transfer matrix $t(u)$ with respect to $u$, we arrive at
\begin{equation}
t(u)=t^{(1)}e^{(N-1)u}+t^{(2)}e^{(N-3)u}+\cdots+t^{(N)}e^{-(N-1)u}. \label{Expansion}
\end{equation}
Here all the expansion coefficients commute with each other and can be regarded as the conserved quantities of the system.
The Hamiltonian (\ref{Ham}) is chosen as the first order derivative of the logarithm of the transfer matrix, namely,
\begin{equation}
H=-2 \sinh \eta  \frac{\partial \ln t(u)}{\partial u}\big|_{u=0, \theta_j=0} + N \cosh \eta. \label{hai11}
\end{equation}

\section{Exact solution at the point of $\eta=\frac{i\pi}{3}$}\label{Interaction-case}
\setcounter{equation}{0}

General spectrum of the Hamiltonian (\ref{Ham}) is obtained by constructing an extended $T-Q$ ansatz \cite{16}.
From the properties (\ref{quasi-}) of $R$-matrix, we know that the transfer matrix (\ref{transfer_matrix}) satisfies
the operator identities
\begin{eqnarray}
&&t(u+i\pi)=(-1)^{N-1}t(u),\label{Quasi-transfer-1}\\
&&t(\theta_j)t(\theta_j-\eta)=-a(\theta_j)\,d(\theta_j-\eta)\times {\rm id},\quad j=1,\ldots, N,\label{Main-identity-1}
\end{eqnarray}
where the functions $a(u)$ and $d(u)$ are given by
\begin{eqnarray}
&&a(u)=\prod_{l=1}^N
\frac{\sinh(u-\theta_l+\eta)} {\sinh\eta},\quad d(u)=a(u-\eta).\label{a-d-functions}
\end{eqnarray}

The commutativity of the transfer matrices with different spectrum
implies that they have  common eigenstates. Let $|\Psi\rangle$ be an
eigenstate of $t(u)$, which does not depend upon $u$,  with
the eigenvalue $\Lambda(u)$, i.e.,
\begin{eqnarray}
t(u)|\Psi\rangle=\Lambda(u)|\Psi\rangle.
\end{eqnarray}
From the definition (\ref{transfer_matrix}) of the transfer matrix and Eqs.(\ref{Quasi-transfer-1})-(\ref{Main-identity-1}),
we obtain that the eigenvalue $\Lambda(u)$ should satisfy
\begin{eqnarray}
&&\Lambda(u) \mbox{, as function of $u$, is a trigonometrical polynomial of degree $N-1$},\label{Eign-anal}\\
&&\Lambda(u+i\pi)=(-1)^{N-1}\Lambda(u),\label{Eigen-Periodic}\\
&&\Lambda(\theta_j)\,\Lambda(\theta_j-\eta)=- a(\theta_j) \,d(\theta_j-\eta),\quad j=1,\ldots, N.\label{Eigen-id}
\end{eqnarray}

Some remarks are in order.
It is shown that the above relations (\ref{Eign-anal})-(\ref{Eigen-id}) allow us to determine the spectrum $\Lambda(u)$ of the transfer matrix $t(u)$, which
is given in terms of the inhomogeneous $T-Q$ relation for a generic $\eta$ \cite{16}.
However, the resulting inhomogeneous $T-Q$ leads to an inhomogeneous BAEs, which is hard to be study if the system size is very large (especially in the thermodynamical limit).
Instead of studying the associated Bethe roots, here we shall consider zero points of eigenvalue $\Lambda(u)$.

Based on the expansion expression (\ref{Expansion}) or the property (\ref{Eign-anal}), we express the eigenvalue
$\Lambda(u)$ in terms of its $N-1$ zero points $\{z_j|j=1,\cdots,N-1\}$ and an overall coefficient
constant $\Lambda_0$ as
\begin{eqnarray}
\Lambda(u)=\Lambda_0\,\prod_{j=1}^{N-1}\,\sinh (u-z_j),\label{Zero-points}
\end{eqnarray}
where the $N$ parameters $\{z_j|j=1,\cdots, N-1\}$ and $\Lambda_0$ can be determined completely by the $N$ equations (\ref{Eigen-id}).

Because $\Lambda(u)$ is the eigenvalue of the transfer matrix $t(u)$, the zero points $\{z_j\}$ must satisfy some constraints, which are
the BAEs of zero points. In this paper, we focus on the case of $\eta=\frac{i\pi}{3}$. Let us introduce a $3N-3$ degree trigonometric polynomial $F_3(u)$
\begin{eqnarray}
F_3(u)=\Lambda(u)\,\Lambda(u-\eta)\,\Lambda(u-2\eta),\label{F3-function}
\end{eqnarray}
which enjoys the properties
\begin{eqnarray}
&& F_3(u+\eta)=(-1)^{N-1}\,F_3(u),\label{quasi-periodic-2}\\
&& F_3(u)=F_3^{(1)}e^{(3N-3)u}+F_3^{(2)}e^{(3N-5)u}+\cdots+F_3^{(3N-2)}e^{-(3N-3)u},\label{quasi-periodic-2-2}\\
&& F_3(\theta_j)=-a(\theta_j)d(\theta_j-\eta)\,\Lambda(\theta_j-2\eta),\quad j=1,\cdots,N, \label{Property-1}\\
&& F_3(\theta_j+\eta)=-a(\theta_j)d(\theta_j-\eta)\,\Lambda(\theta_j+\eta),\quad j=1,\cdots,N, \label{Property-2}\\
&& F_3(\theta_j+2\eta)=(-1)^Na(\theta_j)d(\theta_j-\eta)\,\Lambda(\theta_j+\eta),\quad j=1,\cdots,N. \label{Property-3}
\end{eqnarray}
The above relations (\ref{quasi-periodic-2})-(\ref{Property-3}) can uniquely determine the $3N-3$ trigonometric polynomial $F_3(u)$, and the result is that
the eigenvalue $\Lambda(u)$ should satisfy
\begin{eqnarray}
&&\Lambda(u)\Lambda(u-\eta)\Lambda(u-2\eta)=-a(u)d(u-\eta)\Lambda(u-2\eta)-a(u-\eta)d(u-2\eta)\Lambda(u)\nonumber\\
&&\hspace{4.8cm}+(-1)^Na(u+\eta)d(u)\Lambda(u-\eta).\label{Main-relation-2}
\end{eqnarray}
Substituting the parametrization (\ref{Zero-points}) into Eq.(\ref{Main-relation-2}), we conclude that the zero points $\{z_j|j=1,\cdots,N-1\}$ of the
eigenvalue $\Lambda(u)$ must satisfy the BAEs \footnote{It is remarked that only in the case of $\eta=\frac{i\pi}{2}\,\, {\rm or}\,\,\frac{i\pi}{3}$ the roots of the transfer matrix may satisfy a homogeneous type BAEs like (\ref{BAE-2}). }
\begin{eqnarray}
(-1)^N\frac{d(z_j)}{a(z_j)}=\prod_{l=1}^{N}\frac{\sinh(z_j-\theta_l)}{\sinh(z_j-\theta_l-2\eta)}=
\prod_{k\neq j}^{N-1}\frac{\sinh(z_j-z_k+\eta)}{\sinh(z_j-z_k-\eta)},\;\; j=1,\cdots,N-1.\label{BAE-2}
\end{eqnarray}
The energy of the Hamiltonian (\ref{Ham}) can be expressed in terms of the zero points $\{z_j\}$ as
\begin{eqnarray}
E=2\sinh\eta\sum_{j=1}^{N-1}\coth z_j +N\cosh\eta. \label{BAqE-2}
\end{eqnarray}

Now, we check above analytical results numerically. We first solve the BAEs (\ref{BAE-2}) with $\eta=\frac{i\pi}{3}$
and obtain the values of zero points. Substituting these values into (\ref{BAqE-2}),
we obtain the eigenenergies of the Hamiltonian (\ref{Ham}). The results are listed in table \ref{table:BAEs5}. After
that, we numerically diagonalize the Hamiltonian (\ref{Ham}) with same model parameter. We find
that the eigenvalues obtained by solving the BAEs are exactly the same as those obtained by
the exact diagonalization. Therefore, the expression (\ref{BAqE-2}) gives the complete spectrum of the
system.

\begin{table}[tph]
	\centering 
    \caption{The energy spectrum of the Hamiltonian (\ref{Ham}) with $N=6$ and $\eta=\frac{i\pi}{3}$.
    Here $\{\lambda_j=z_j+\frac\eta 2|j=1,\cdots, 5\}$ are the shifted zero points, $E_n$ is the eigenenergy and $n$ is the energy level.
    The energy $E_n$ calculated from the BAEs (\ref{BAE-2}) is exactly the same as that obtained from the exact diagonalization of the Hamiltonian (\ref{Ham}).}
	\begin{scriptsize}
		\begin{tabular}{ccccccc} \bottomrule
			$\lambda_{1}$ & $\lambda_{2}$ & $\lambda_{3}$ & $\lambda_{4}$ & $\lambda_5$ & $E_{n}$ & $n$  \\  \bottomrule
			$0$ & $0.6969$ & $-0.6969$ & $0.2664$ & $-0.2664$ & $-6.4785$ & $1$  \\
			$-0.6700$ &  $0.0310$ & $0.3300$ & $-0.2415$ & $1.0477-1.5708i$ &$-5.1280$ & $2$  \\
			$0.6700$ & $-0.0310$ & $-0.3300$ & $0.2415$ & $-1.0477+1.5708i$ & $-5.1280$ & $3$  \\
			$-0.0804$ & $-0.4734$ & $0.2076$ & $0.6357$ & $-0.3695-1.5708i$ & $-4.2182$ & $4$  \\
			$0.0804$ & $0.4734$ & $-0.2076$ & $-0.6357$ & $0.3695+1.5708i$ & $-4.2182$ & $5$  \\
			$0.1528$ & $-0.1528$ & $-0.5836$ & $0.5836$ & $-1.5708i$ & $-3.8607$ & $6$  \\
			$-0.3047$ & $0.3047$ & $0$ & $1.0124-1.5708i$ & $-1.0124+1.5708i$ & $-3.6496$ & $7$  \\
			$-0.2687$ & $0.4515$ & $0.0516$ & $0.3238-1.5708i$ & $-0.9667+1.5708i$ & $-2.6398$ & $8$  \\
			$0.2687$ & $-0.4515$ & $-0.0516$ & $-0.3238+1.5708i$ & $0.9667+1.5708i$ & $-2.6398$ & $9$  \\
			$0.1301$ & $-0.2030$ & $0.5604$ & $-0.0819-1.5708i$ & $-0.8975+1.5708i$ & $-2.2241$ & $10$  \\
			$-0.1301$ & $0.2030$ & $-0.5604$ & $0.0819-1.5708i$ & $0.8975+1.5708i$ & $-2.2241$ & $11$  \\
			$0.6203$ & $0.1974$ & $-0.0862$ & $-0.4537+1.0526i$ & $-0.4537-1.0526i$ & $-2.0000$ & $12$  \\
			$-0.6203$ & $-0.1974$ & $0.0862$ & $0.4537+1.0526i$ & $0.4537-1.0526i$ & $-2.0000$ & $13$  \\
			$0$ & $-0.4175$ & $0.4175$ & $0.2632+1.5708i$ & $-0.2632+1.5708i$ & $-1.6234$ & $14$  \\
			$0.1817$ & $-0.1817$ & $1.5708i$ & $-0.8587-1.5708i$ & $0.8587+1.5708i$ & $-0.4043$ & $15$  \\
			$0.0541$ & $-0.2606$ & $0.4254+1.0539i$ & $0.4254-1.0539i$ & $-0.9414-1.5708i$ & $-0.2277$ & $16$  \\
			$-0.0541$ & $0.2606$ & $-0.4254+1.0539i$ & $-0.4254-1.0539i$ & $0.9414+1.5708i$ & $-0.2277$ & $17$  \\
			$-0.5749$ & $-0.1613$ & $0.5231$ & $0.1269+1.0479i$ & $0.1269-1.0479i$ & $0.5222$ & $18$  \\
			$0.5749$ & $0.1613$ & $-0.5231$ & $-0.1269+1.0479i$ & $-0.1269-1.0479i$ & $0.5222$ & $19$  \\
			$0.0067$ & $0.4202$ & $0.2161-1.5708i$ & $-0.3803+1.0558i$ & $-0.3803-1.0558i$ & $1.0892$ & $20$  \\
			$-0.0067$ & $-0.4202$ & $-0.2161-1.5708i$ & $0.3803+1.0558i$ &$0.3803-1.0558i$ & $1.0892$ & $21$  \\
			$-0.5380$ & $-0.1223$ & $0.2012+1.0510i$ & $0.2012-1.0510i$ & $0.6784-1.5708i$ & $2.0000$ & $22$  \\
			$0.5380$ & $0.1223$ & $-0.2012+1.0510i$ & $-0.2012-1.0510i$ & $-0.6784+1.5708i$ & $2.0000$ & $23$  \\
			$0.4900$ & $-0.2257$ & $0.0925-1.0486i$ & $0.0925+1.0486i$ & $-0.8593-1.5708i$ & $2.3827$ & $24$  \\
			$-0.4900$ & $0.2257$ & $-0.0925-1.0486i$ & $-0.0925+1.0486i$ & $0.8593-1.5708i$ & $2.3827$ & $25$  \\
			$0.4146$ & $-0.4146$ & $1.5708i$ & $+1.0499i$ & $-1.0499i$ & $3.5137$ & $26$  \\
			$0$ & $0.3767-1.0590i$ & $0.3767+1.0590i$ & $-0.3767-1.0590i$ & $-0.3767+1.0590i$ & $3.7515$ & $27$  \\
			$0.1845$ & $-0.1721+1.0544i$ & $-0.1721-1.0544i$ & $-0.5975-1.5708i$ & $0.7995-1.5708i$ & $4.1565$ & $28$  \\
			$-0.1845$ & $0.1721-1.0544i$ & $0.1721+1.0544i$ & $0.5975-1.5708i$ & $-0.7995-1.5708i$ & $4.1565$ & $29$  \\
			$-0.0279-1.0558i$ & $-0.0279+1.0558i$ & $0.3317+1.0832i$ & $0.3317-1.0832i$ & $-0.4288$ & $6.2874$ & $30$  \\
			$0.4288$ & $0.0279+1.0558i$ & $0.0279-1.0558i$ & $-0.3317+1.0832i$ & $-0.3317-1.0832i$ & $6.2874$ & $31$  \\
			$-1.5708i$ & $0.1962+1.1268i$ & $0.1962-1.1268i$ & $-0.1962-1.1268i$ & $-0.1962+1.1268i$ & $8.7513$ & $32$  \\
			\bottomrule \end{tabular}
	\end{scriptsize}\label{table:BAEs5}
\end{table}

\section{Ground state energy}\label{the gs}
\setcounter{equation}{0}

\begin{figure}
	\noindent\centering
	\includegraphics[width=9cm]{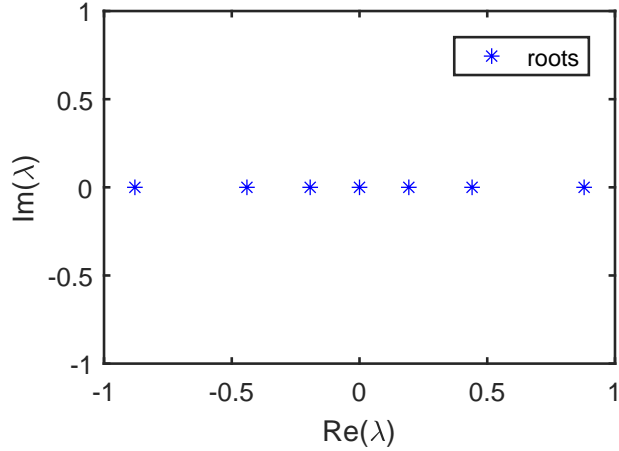}
	\caption{The patterns of zero points at the ground state of the system (\ref{Ham}). Here $N=8$ and $\eta = \frac{i}{3} \pi$.}
	\label{ground}
\end{figure}

At the ground state, all the zero points are real. The pattern of zero points with $N=6$ is shown in Fig.\ref{ground}.
Because the even $N$ and the odd $N$ give the same thermodynamic limit, we set $N$ as an even number in the following.
The zero points at the ground state should satisfy the BAEs
\begin{eqnarray}\label{ZES}
\left[\frac{\sinh(\lambda_j-\frac{i}{6}\pi)}
{\sinh(\lambda_j+\frac{i}{6}\pi)}\right]^N
= -\prod_{k=1}^{N-1}
\frac{\sinh(\lambda_j-\lambda_k+\frac{i}{3}\pi)}
{\sinh(\lambda_j-\lambda_k-\frac{i}{3}\pi)}, \quad j=1, \cdots, N-1,
\end{eqnarray}
where $\lambda_j=z_j+\frac\eta 2$.
Taking the logarithm of Eq.(\ref{ZES}), we have
\begin{eqnarray}
\label{BAEf01}
-Ni\ln
\frac{\sinh(\frac{i}{6}\pi-\lambda_j)}
{\sinh(\frac{i}{6}\pi+\lambda_j)}
= 2\pi I_j+
i\ln \frac{\sinh[\frac{i}{3}\pi-(\lambda_j-\lambda_k)]}
{\sinh[\frac{i}{3}\pi+(\lambda_j-\lambda_k)]},\quad j=1, \cdots, N-1,
\end{eqnarray}
where $I_j$ is the quantum number. At the ground state, the set of quantum numbers reads
\begin{equation}\label{1}
\{I_j\}=\Big\{-\frac{N}{2}+1, -\frac{N}{2}+2,\cdots, \frac{N}{2}-2,\frac{N}{2}-1\Big\}.
\end{equation}
We note that there are $N-2$ quantum numbers in the set.
For simplicity, we define
\begin{eqnarray}
\theta_m(\lambda) =-i\ln
\frac{\sinh(\frac{i \pi m}{6}-\lambda_j)}
{\sinh(\frac{i \pi m}{6}+\lambda_j)}.
\end{eqnarray}
Then Eq.\eqref{BAEf01} can be rewritten as
\begin{eqnarray}
\theta_1(\lambda_j)= \frac{2\pi I_j}{N}
-\frac{1}{N}\sum_{k=1}^{N-1}\theta_2(\lambda_j-\lambda_k). \label{ai2}
\end{eqnarray}

In the thermodynamic limit, the distribution of zero points $\{\lambda_j | j=1, \cdots, N-1\}$ tends to continuous. Thus
we denote the zero point $\lambda_j$ by a continuous variable $\lambda$.
Define the counting function
\begin{equation}\label{BAEf02}
Z(\lambda)=\frac{1}{2\pi}\left[\theta_1(\lambda)+
\frac{1}{N}\sum_{k=1}^{N-1}\theta_2(\lambda-\lambda_k)\right].
\end{equation}
The derivative of counting function $Z(\lambda)$ gives
\begin{equation}
\frac{dZ(\lambda)}{d\lambda}\equiv
\rho_g(\lambda)+\rho_g^h(\lambda),
\end{equation}
where $\rho_g(\lambda)$ and $\rho_g^h(\lambda)$ are the density of particles and that of holes, respectively.
Taking the derivative of BAEs \eqref{ai2} in the thermodynamic limit with respect to $\lambda$,
we obtain
\begin{eqnarray}
\rho_g(\lambda)+\rho_g^h(\lambda) =
a_1(\lambda)+
\int_{-\infty}^{\infty}a_2(\lambda-\mu)\rho_g(\mu)d\mu,
\label{BAEf03}
\end{eqnarray}
where the function $a_m(\lambda)$ is given by
\begin{eqnarray}
a_m(\lambda) =
\frac{1}{2\pi}\frac{d\theta_m(\lambda)}{d\lambda}
=\frac{1}{\pi}
\frac{\sin(m\gamma)}{\cosh(2\lambda)-\cos(m\gamma)}, \quad \gamma=\frac{\pi}{3},
\end{eqnarray}
and the density of holes is
\begin{eqnarray}
\rho_g^h(\lambda) = \frac{1}{3N}\delta(\lambda-\lambda_0^h)
+\frac{1}{3N}\delta(\lambda+\lambda_0^h).\label{ai3}
\end{eqnarray}
We shall note that the quantum numbers $\{I_j\}$ in Eq.\eqref{1} are the continuous integers, i.e., $I_j-I_{j-1}=1$, which means that there is no holes in the bulk.
Only the boundary holes associated with the twisted bond (spin-exchanging interaction with nearest neighbor) are allowed. Due to the facts that there are $N-2$ quantum numbers in Eq.\eqref{1} and the energy should be real, we conclude that there are two symmetric holes at the positions of
$\pm\lambda_0^h$ with the coefficient $\frac1{3N}$.
The density $\rho_g(\lambda)$ should satisfy
\begin{equation}\label{2}
\int_{-\infty}^{\infty}\rho_g(\lambda)d\lambda=\frac{N-1}{N},
\end{equation}
because that the number of zero points is $N-1$.
The integral equation \eqref{BAEf03} can be solved by the Fourier transformation defined
for an arbitrary function $F(\lambda)$ as
\begin{eqnarray}
\tilde{F}(w) = \int_{-\infty}^{\infty}e^{iw\lambda}F(\lambda)d\lambda,\quad
F(\lambda) = \frac{1}{2\pi}
\int_{-\infty}^{\infty}e^{-iw\lambda}\tilde{F}(w)dw.\label{fly02}
\end{eqnarray}
Taking the Fourier transform of Eq.\eqref{BAEf03}, we obtain
\begin{equation}
\tilde{\rho}_g(w)=\frac1{1-\tilde{a}_2(w)}
\Big[\tilde{a}_1(w)-\frac{e^{iw\lambda_0^h}}{3N}
-\frac{e^{-iw\lambda_0^h}}{3N}\Big],\label{ai4}
\end{equation}
where the Fourier transform of function $a_m(\lambda)$ is
\begin{eqnarray}
\tilde{a}_m(w) = \frac{\sinh(\pi w/2-\delta_m\pi w)}
{\sinh(\pi w/2)}, \quad \delta_m =\frac{m\gamma}{2\pi}.
\end{eqnarray}
The inverse Fourier transform of Eq.(\ref{ai4}) gives the solution of integral equation \eqref{BAEf03} as
\begin{eqnarray}\label{rho1}
&&\rho_g(\lambda)
=\frac{3i}{4\pi}\Big[{\rm  csch}\big(\frac{3}{2}\lambda+\frac{i\pi}{4}\big) -{\rm csch}\big(\frac{3}{2}\lambda-\frac{i\pi}{4}\big)\Big]
-\frac{1}{3N}\Big\{\delta(\lambda-\lambda_0^h)+\delta(\lambda+\lambda_0^h) \nonumber \\[6pt]
&&\qquad \qquad +\frac{3}{4\pi}{\rm sech}\big[\frac{3}{2}(\lambda-\lambda_0^h)\big] +\frac{3}{4\pi}{\rm sech}\big[\frac{3}{2}(\lambda+\lambda_0^h)\big]\Big\}.
\end{eqnarray}
Thus the ground state energy is
\begin{eqnarray}
&&E_g=2N\sinh\big(\frac{i}{3}\pi\big)\int_{-\infty}^\infty\coth\big(\lambda-\frac{i}{6}\pi\big)
\rho_g(\lambda)d\lambda+N\cosh(\frac{i}{3}\pi)\nonumber\\
&&\quad\;\; =\frac{3-3\sqrt{3}}{2}N+\bigtriangleup(\lambda^h_0)+\bigtriangleup(-\lambda^h_0),
\end{eqnarray}
where $\bigtriangleup(\lambda_0^h)$ is the contribution of hole $\lambda_0^h$ with the form of
\begin{eqnarray}\label{delta}
\bigtriangleup(\lambda_0^h)=\frac{\sqrt{3}}{4}\Big[{\rm sech}\big(\frac{3}{2}\lambda_0^h +\frac{i\pi}{4}\big)+{\rm sech}\big(\frac{3}{2}\lambda_0^h-\frac{i\pi}{4}\big)\Big] +\frac{\sqrt{3}}{2}i\tanh(3\lambda_0^h).
\end{eqnarray}
In the thermodynamic limit, the holes should be located at the infinity in the real axis, i.e., $\lambda_0^h\rightarrow \infty$, to minimize the energy.
Then we obtain the density of ground state energy
\begin{eqnarray}
e_g=\frac{3-3\sqrt{3}}{2},
\end{eqnarray}
which is equal to the density of ground state energy of the system with the periodic boundary condition given in reference \cite{22}.
This result is reasonable. There is only one bond is twisted and its contribution to the total energy of the system is infinitesimal when the system size tends to infinity.

\section{Elementary excitations}\label{the ee}
\setcounter{equation}{0}

\subsection{Elementary excitation of type I }

\begin{figure}
	\noindent\centering
	\includegraphics[width=9cm]{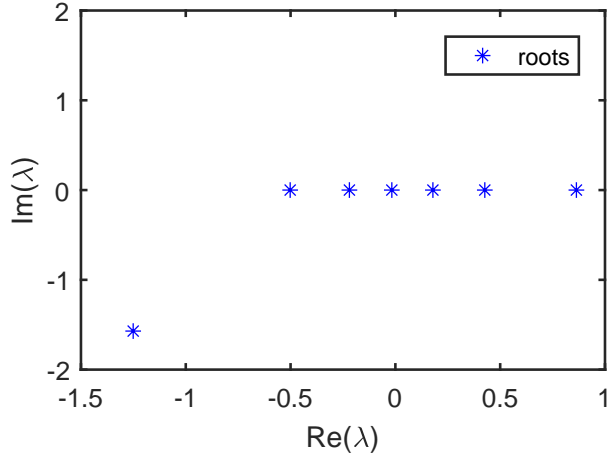}
	\caption{The pattern of zero points at the first kind of elementary excitation. Here $N=8$ and $\eta = \frac{i}{3} \pi$.}
	\label{e1}
\end{figure}

\begin{figure}
	\noindent\centering
	\includegraphics[width=9cm]{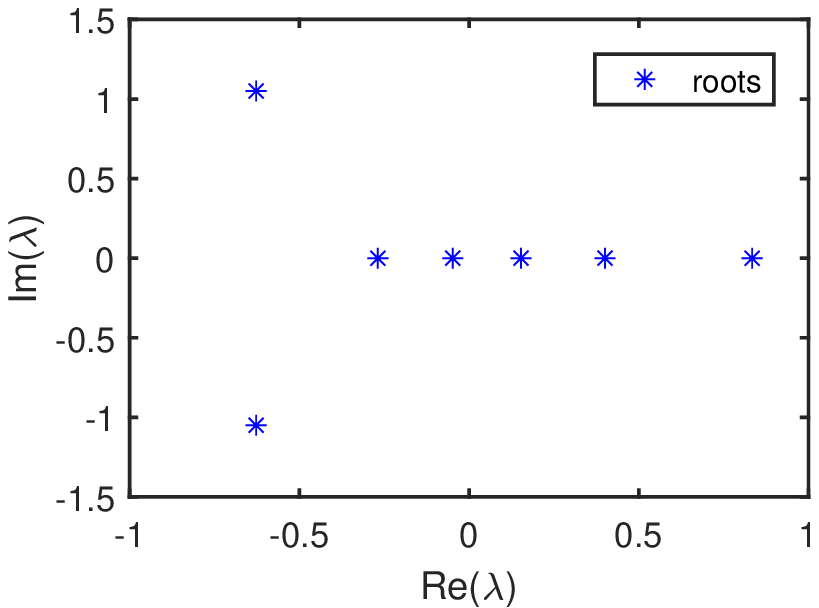}
	\caption{The pattern of zero points at the second kind of elementary excitation. Here $N=8$ and $\eta = \frac{i}{3} \pi$.}
	\label{e2}
\end{figure}

In this section, we consider the elementary excitations in the system (\ref{Ham}). There are two kinds of elementary excitations.
The first case is that one real zero point is excited to the line with fixed imaginary part $-i\frac{\pi}{2}$, while the rest zero points remain real.
The corresponding pattern of zero points distribution with $N=8$ is shown in Fig.\ref{e1}.
The second case is that two zero points are excited to the lines with fixed imaginary parts $\pm\eta$, while the rest zero points remain real.
We shall note that these two excited zero points form the $2$-string with the length of $2|\eta|$. The corresponding pattern of zero points distribution with $N=8$ is shown in Fig.\ref{e2}.

In the first case, denote the excited zero point as $\lambda_{N-1}=\alpha-i\frac{\pi}{2}$, where $\alpha$ is real.
Then the BAEs \eqref{ZES} reads
\begin{eqnarray}
&&\left[\frac{\sinh(\lambda_j-\frac{i}{6}\pi)}
{\sinh(\lambda_j+\frac{i}{6}\pi)}\right]^N
= -\prod_{k=1}^{N-2}
\frac{\sinh(\lambda_j-\lambda_k+\frac{i}{3}\pi)}
{\sinh(\lambda_j-\lambda_k-\frac{i}{3}\pi)}
\frac{\cosh(\lambda_j-\alpha+\frac{i}{3}\pi)}
{\cosh(\lambda_j-\alpha-\frac{i}{3}\pi)}, \;\; j=1, \cdots, N-2, \nonumber \\ \label{ZESE1_1}
&&\left[\frac{\cosh(\alpha-\frac{i}{6}\pi)}
{\cosh(\alpha+\frac{i}{6}\pi)}\right]^N
= \prod_{k=1}^{N-2}
\frac{\cosh(\alpha-\lambda_k+\frac{i}{3}\pi)}
{\cosh(\alpha-\lambda_k-\frac{i}{3}\pi)}.
\end{eqnarray}
Taking the logarithm of Eq.\eqref{ZESE1_1}, we have
\begin{eqnarray}
&&\theta_1(\lambda_j) = \frac{2\pi I_j}{N}
-\frac{1}{N}\sum_{k=1}^{N-2}\theta_2(\lambda_j-\lambda_k)
+\frac{1}{N}\theta_1(\lambda_j-\alpha),\nonumber \\
&&\theta_2(\alpha) = \frac{2\pi J}{N}
-\frac{1}{N}\sum_{k=1}^{N-2}\theta_1(\alpha-\lambda_k), \label{BAEf24}
\end{eqnarray}
where the sets of quantum numbers are
\begin{eqnarray}
&& \{I_j\} = \Big\{-\frac{N-1}{2}+1,-\frac{N-1}{2}+2,\cdots,\frac{N-1}{2}-2,\frac{N-1}{2}-1\Big\}, \label{EE1I}\\
&&J \in \Big\{-\frac{N}{2}+1,-\frac{N}{2}+2,\cdots,\frac{N}{2}-2,\frac{N}{2}-1\Big\}.\label{EE1J}
\end{eqnarray}
We shall note that the quantum number $J$ takes one value in the set given by Eq.\eqref{EE1J}.
In the thermodynamic limit, $\lambda_j$ becomes a continuous variable $\lambda$.
Taking the derivative of equation \eqref{BAEf24} with respect to $\lambda$,
we obtain
\begin{eqnarray}
&&\rho_{1}(\lambda)+\rho_1^h(\lambda)=
a_1(\lambda)+
\int_{-\infty}^{\infty}a_2(\lambda-\mu)\rho_{1}(\mu)d\mu
-\frac{1}{N}a_1(\lambda-\alpha), \label{BAEf25-1} \\
&&\rho_{1}^h(\lambda)=\frac{1}{3N}\delta(\lambda-\lambda_0^h)+\frac{1}{3N}\delta(\lambda+\lambda_0^h).\label{BAEf25}
\end{eqnarray}
From Eq.\eqref{EE1I}, we see that the quantum numbers $\{I_j\}$ are continuous.
Thus the structure of holes is the same as that at the ground state. From Eq.\eqref{EE1J}, we know that the
excited zero point contributes the last term in \eqref{BAEf25-1}. This conclusion can also be obtained
by comparing Eqs.\eqref{BAEf03} and \eqref{BAEf25-1}.
The density of zero points should satisfy the constraint
\begin{eqnarray}
\int_{-\infty}^{\infty}\rho_1(\lambda)d\lambda=\frac{N-2}{N}.
\end{eqnarray}
With the help of Fourier transform, the solution of integral equation \eqref{BAEf25-1} is
\begin{eqnarray}
\tilde{\rho}_1(w)=
\frac{1}{1-\tilde{a}_2(w)}
\Big\{\tilde{a}_1(w)-\frac{e^{iw\alpha}\tilde{a}_1(w)}{N}
-\frac{e^{-iw\lambda_0^h}}{3N}
-\frac{e^{-iw\lambda_0^h}}{3N}\Big\}.\label{drhwo1w}
\end{eqnarray}
Define the density difference between the excited and the ground state as
\begin{eqnarray}
\delta\tilde{\rho}_1(w) \equiv \tilde{\rho}_1(w)-\tilde{\rho}_g(w)=-
\frac{1}{N[1-\tilde{a}_2(w)]} e^{iw\alpha}\tilde{a}_1(w).\label{drho1w}
\end{eqnarray}
The inverse Fourier transform of Eq.\eqref{drho1w} gives the density changing at the excited state as
\begin{eqnarray}\label{drho1}
\delta\rho_{1}(\lambda)=
-\frac{3i}{4\pi N}\Big[{\rm csch}\big(\frac{3}{2}\lambda-\frac{3}{2}\alpha+\frac{i\pi}{4}\big) -{\rm csch}\big(\frac{3}{2}\lambda-\frac{3}{2}\alpha-\frac{i\pi}{4}\big)\Big].
\end{eqnarray}
Thus the elementary excitation carries the energy
\begin{eqnarray}
&&\delta e_1=2N\sinh\big(\frac{i}{3}\pi\big)\int_{-\infty}^\infty\coth\big(\lambda-\frac{i}{6}\pi\big)
\delta\rho_1(\lambda)d\lambda+2\sinh\big(\frac{i}{3}\pi\big)\coth\big(\alpha-\frac{2i}{3}\pi\big)
\nonumber \\
&&\quad\;\; =\frac{3\sqrt{3}}{2}{\rm sech}\big(\frac{3\alpha}{2}\big).
\end{eqnarray}

\subsection{Elementary excitation of type II}

Now we consider the second kind of elementary excitation. Because two zero points form the $2$-string,
we set $\lambda_{N-2}=\alpha-i\frac{\pi}{3}$ and $\lambda_{N-1}=\alpha+i\frac{\pi}{3}$, where $\alpha$ is real.
The BAEs (\ref{ZES}) in the case are
\begin{eqnarray}
&&\left[\frac{\sinh(\lambda_j-\frac{i}{6}\pi)}
{\sinh(\lambda_j+\frac{i}{6}\pi)}\right]^N
= -\prod_{k=1}^{N-3}
\frac{\sinh(\lambda_j-\lambda_k+\frac{i}{3}\pi)}
{\sinh(\lambda_j-\lambda_k-\frac{i}{3}\pi)}
\frac{\sinh(\lambda_j-\alpha+\frac{2i}{3}\pi)}
{\sinh(\lambda_j-\alpha-\frac{2i}{3}\pi)}, \nonumber \\
&&\qquad \;\; j=1,\cdots, N-3, \label{ZES2_0} \\
&&\left[\frac{\sinh(\alpha+\frac{i}{6}\pi)}
{\sinh(\alpha-\frac{i}{6}\pi)}\right]^N
= \prod_{k=1}^{N-3}
\frac{\sinh(\alpha-\lambda_k+\frac{2i}{3}\pi)}
{\sinh(\alpha-\lambda_k-\frac{2i}{3}\pi)}.\label{ZES2_1}
\end{eqnarray}
We shall note that substituting $\lambda_{N-2}=\alpha-i\frac{\pi}{3}$ and $\lambda_{N-1}=\alpha+i\frac{\pi}{3}$ into BAEs (\ref{ZES})
and taking the product of these two equations, we arrive at Eq.(\ref{ZES2_1}).
Taking the logarithm of equations \eqref{ZES2_0} and \eqref{ZES2_1}, we have
\begin{eqnarray}
&&\theta_1(\lambda_j) = \frac{2\pi I_j}{N}
-\frac{1}{N}\sum_{k=1}^{N-3}\theta_2(\lambda_j-\lambda_k)
+\frac{1}{N}\theta_2(\lambda_j-\alpha),\label{BAEf51_0} \\
&&\theta_1(\alpha) = \frac{2\pi J}{N}
-\frac{1}{N}\sum_{k=1}^{N-3}\theta_2(\alpha-\lambda_k),\label{BAEf51}
\end{eqnarray}
where the sets of quantum numbers are given by
\begin{eqnarray}
&&\{I_j\} = \Big\{-\frac{N-1}{2}+1,-\frac{N-1}{2}+2,\cdots,\frac{N-1}{2}-j-1, \nonumber \\
&&\qquad\qquad \frac{N-1}{2}-j+1, \cdots,\frac{N-1}{2}-2,\frac{N-1}{2}-1\Big\}, \label{BAE1f51} \\
&& J= \frac{N-1}{2}-j, \quad j=1,\cdots,N-2.\label{BAE1f51-1}
\end{eqnarray}
We see that the quantum numbers $\{I_j\}$ in Eq.\eqref{BAE1f51} are discontinuous at the point of $j$. Thus there exists one bulk hole, which is induced by the 2-string excitation.
Taking the derivative of equations \eqref{BAEf51_0} and \eqref{BAEf51} with respect to $\lambda$, we obtain
\begin{eqnarray}
&&\rho_{2}(\lambda)+\rho_{2}^h(\lambda)=
a_1(\lambda)+
\int_{-\infty}^{\infty}a_2(\lambda-\mu)\rho_{2}(\mu)d\mu
+\frac{1}{N}a_4(\lambda-\alpha), \label{BeAE1f51-0} \\
&&\rho_{2}^h(\lambda)= \frac{1}{N}\delta(\lambda-\lambda_2^h) +\frac{1}{3N}\delta(\lambda-\lambda_0^h)+\frac{1}{3N}\delta(\lambda+\lambda_0^h),\label{BeAE1f51-1}
\end{eqnarray}
where the $\lambda_2^h$ is the position of hole induced by the 2-string.
The density of zero roots should satisfy the constraint
\begin{eqnarray}
\int_{-\infty}^{\infty}\rho_{2}(\lambda)d\lambda=\frac{N-3}{N}.
\end{eqnarray}
With the help of Fourier transform, we obtain the solution of integral equations \eqref{BeAE1f51-0} and \eqref{BeAE1f51-1} as
\begin{eqnarray}\label{rho2w}
\tilde{\rho}_{2}(w)=
\frac{1}{1-\tilde{a}_2(w)}\Big\{
\tilde{a}_1(w)+\frac{1}{N}e^{iw\alpha}\tilde{a}_4(w)
-\frac{e^{iw\lambda_2^h}}{N}
-\frac{e^{-iw\lambda_0^h}}{3N}
-\frac{e^{iw\lambda_0^h}}{3N}\Big\}.
\end{eqnarray}
The density different between the excited state and the ground state reads
\begin{eqnarray}
\delta\tilde{\rho}_2(w)\equiv\tilde{\rho}_2(w)-\tilde{\rho}_g(w)=
\frac{1}{N[1-\tilde{a}_2(w)]}
\Big\{e^{iw\alpha}\tilde{a}_4(w)-e^{iw\lambda_2^h}\Big\}.\label{drho2w}
\end{eqnarray}
Thus, we obtain the density changing at the excited state
\begin{eqnarray}\label{drho2}
\delta\rho_{2}(\lambda)=-\frac{3}{4\pi N}\left[{\rm sech}(\frac{3}{2}\lambda-\frac{3}{2}\alpha)+{\rm sech}(\frac{3}{2}\lambda-\frac{3}{2}\lambda_2^h)\right]
-\frac{1}{N}\delta(\lambda-\lambda_2^h).
\end{eqnarray}
The elementary excitation carries the energy
\begin{eqnarray}
&&\delta e_2=2N\sinh\big(\frac{i}{3}\pi\big)\int_{-\infty}^\infty\coth\big(\lambda-\frac{i}{6}\pi\big)
\delta\rho_2(\lambda)d\lambda+2\sinh\big(\frac{i}{3}\pi\big)\coth\big(\alpha-\frac{i}{2}\pi\big)\nonumber\\
&&\quad \quad \;\;+2\sinh\big(\frac{i}{3}\pi\big)\coth\big(\alpha+\frac{i}{6}\pi\big) \nonumber \\
&& \quad \;\; = 3\triangle(\alpha)+3\triangle(-\lambda_2^h). \label{drwho2w}
\end{eqnarray}
From Eq.\eqref{BAE1f51}, we see that the hole in the set of quantum number $\{I_j\}$ is $\frac{N-1}{2}-j$, which is equal to
the quantum number $J$ \eqref{BAE1f51-1}. Therefore, we have $\lambda_2^h=\alpha$.
This conclusion can also be obtained from the BAEs.
We shall note that the hole $\lambda_2^h$ also is a possible solution of BAEs, which leads to
\begin{eqnarray}
\left[\frac{\sinh(\lambda^h_2-\frac{i}{6}\pi)}
{\sinh(\lambda^h_2+\frac{i}{6}\pi)}\right]^N
=-\prod_{k=1}^{N-3}
\frac{\sinh(\lambda^h_2-\lambda_k+\frac{i}{3}\pi)}
{\sinh(\lambda^h_2-\lambda_k-\frac{i}{3}\pi)}
\frac{\sinh(\lambda^h_2-\alpha+\frac{2i}{3}\pi)}
{\sinh(\lambda^h_2-\alpha-\frac{2i}{3}\pi)}.\label{BAE_hole}
\end{eqnarray}
Substituting $\lambda_2^h=\alpha$ into above equation \eqref{BAE_hole}, we arrive at the BAE \eqref{ZES2_1} which means that
$\lambda_2^h=\alpha$ is a solution of equation \eqref{ZES2_1}.
Then the excited energy \eqref{drwho2w} reads
\begin{eqnarray}
\delta e_2= 3\triangle(\alpha)+3\triangle(-\alpha)= \frac{3\sqrt{6}\cosh(\frac{3\alpha}{2})}{\cosh(3\alpha)}.
\end{eqnarray}

\section{Two-body scattering matrix}
\label{scattering matrix}
\setcounter{equation}{0}

In this section, we study the two-body scattering matrix of the elementary excitations. Due to the existence of two kinds of elementary excitations, there are three kinds of scattering processes.

(i) We consider the scattering of two first kind of elementary excitations.
We set $\lambda_{N-2}=\alpha_1-i\frac{\pi}{2}$ and $\lambda_{N-1}=\alpha_2-i\frac{\pi}{2}$, where both $\alpha_1$ and $\alpha_2$ are real.
The other zero points are real. In this case, the BAEs read
\begin{eqnarray}
&&\left[\frac{\sinh(\lambda_j-\frac{i}{6}\pi)}
{\sinh(\lambda_j+\frac{i}{6}\pi)}\right]^N
= -\prod_{k=1}^{N-3}
\frac{\sinh(\lambda_j-\lambda_k+\frac{i}{3}\pi)}
{\sinh(\lambda_j-\lambda_k-\frac{i}{3}\pi)}
\prod_{l=1}^{2}
\frac{\cosh(\lambda_j-\alpha_l+\frac{i}{3}\pi)}
{\cosh(\lambda_j-\alpha_l-\frac{i}{3}\pi)},\nonumber \\
&&\qquad  j=1,\cdots, N-3, \label{E11_10} \\
&&\left[\frac{\sinh(\frac{i}{3}\pi-\alpha_m)}
{\sinh(\frac{i}{3}\pi+\alpha_m)}\right]^N
=\prod_{k=1}^{N-3}
\frac{\sinh(\frac{i}{6}\pi+\alpha_m-\lambda_k)}
{\sinh(\frac{i}{6}\pi-\alpha_m+\lambda_k)}
\frac{\sinh(\frac{i}{3}\pi-\alpha_m+\alpha_p)}
{\sinh(\frac{i}{3}\pi+\alpha_m-\alpha_p)},\nonumber \\
&&\qquad m=1,2, \quad p=1,2, \quad m\neq p.\label{E11_1}
\end{eqnarray}
Taking the logarithm of Eq.\eqref{E11_10}, we have
\begin{eqnarray}\label{logE11}
\theta_1(\lambda_j) = \frac{2\pi I_j}{N}
-\frac{1}{N}\sum_{k=1}^{N-3}\theta_2(\lambda_j-\lambda_k)
+\frac{1}{N}\theta_1(\lambda_j-\alpha_1)+\frac{1}{N}\theta_1(\lambda_j-\alpha_2).
\end{eqnarray}
In the thermodynamic limit, the integral form of Eq.\eqref{logE11} is
\begin{eqnarray}
&&\hspace{-0.8cm}\rho_{11}(\lambda)+\rho_{11}^h(\lambda)=
a_1(\lambda)+
\int_{-\infty}^{\infty}a_2(\lambda-\mu)\rho_{11}(\mu)d\mu
-\frac{1}{N}a_1(\lambda-\alpha_1)-\frac{1}{N}a_1(\lambda-\alpha_2), \label{logE11_10} \\
&&\hspace{-0.8cm}\rho_{11}^h(\lambda) =\frac{1}{3N}\delta(\lambda-\lambda_0^h)
+\frac{1}{3N}\delta(\lambda+\lambda_0^h).\label{logE11_1}
\end{eqnarray}
The present density of zero points $\rho_{11}(\lambda)$ should satisfy
\begin{eqnarray}
\int_{-\infty}^{\infty}\rho_{11}(\lambda)d\lambda=\frac{N-3}{N}.
\end{eqnarray}
With the help of Fourier transform, we obtain the solution of Eq.(\ref{logE11_10}) as
\begin{eqnarray}\label{rho11w}
&&\tilde{\rho}_{11}(w)=
\frac{1}{1-\tilde{a}_2(w)}
\Big\{\tilde{a}_1(w)-\frac{e^{iw\lambda_0^h}}{3N}
-\frac{e^{iw\lambda_0^h}}{3N}-\frac{1}{N}e^{iw\alpha_1}\tilde{a}_1(w)
-\frac{1}{N}e^{iw\alpha_2}\tilde{a}_1(w)\Big\} \nonumber\\[6pt]
&&\qquad\quad \equiv \tilde{\rho}_g(w)+\delta\tilde{\rho}_1(w)_{\alpha_1}+\delta\tilde{\rho}_1(w)_{\alpha_2},
\end{eqnarray}
where $\delta\tilde{\rho}_1(w)_{\alpha_1}$ and $\delta\tilde{\rho}_1(w)_{\alpha_2}$ are given by~\eqref{drho1w}
with the replacing of $\alpha$ by $\alpha_1$ and $\alpha_2$, respectively.
Thus the density of zero points is
\begin{eqnarray}\label{rho11}
\rho_{11}(\lambda)=\rho_g(\lambda)+\delta\rho_1(\lambda)_{\alpha_1}+\delta\rho_1(\lambda)_{\alpha_2},
\end{eqnarray}
where $\delta\rho_1(\lambda)_{\alpha_1}$ and $\delta\rho_1(\lambda)_{\alpha_2}$ are given by~\eqref{drho1}
with the replacing of $\alpha$ by $\alpha_1$ and $\alpha_2$, respectively.

In the thermodynamic limit, we rewrite the BAEs~\eqref{E11_1} as
\begin{eqnarray}
\left[\frac{\sinh(\frac{i}{3}\pi-\alpha_m)}
{\sinh(\frac{i}{3}\pi+\alpha_m)}\right]^N
=e^{-Ni\int_{-\infty}^{\infty}\theta_1(\alpha_m-\lambda)\rho_{11}(\lambda)d\lambda}
\frac{\sinh[\frac{i}{3}\pi-(\alpha_m-\alpha_p)]}
{\sinh[\frac{i}{3}\pi+(\alpha_m-\alpha_p)]}, m,p=1,2; m\neq p.\label{E11_2}
\end{eqnarray}
Substituting Eq.\eqref{rho11} into~\eqref{E11_2} and taking the integral, we have
\begin{eqnarray}
\left[\frac{i-\sinh(\frac{3}{2}\alpha_m)}{i+\sinh(\frac{3}{2}\alpha_m)}\right]^N =\frac{i-\sinh[\frac{3}{2}(\alpha_m-\alpha_p)]}{i+\sinh[\frac{3}{2}(\alpha_m-\alpha_p)]}, \;\; m,p=1,2, \;\; m\neq p.
\end{eqnarray}
Then the two-body scattering matrix is
\begin{eqnarray}
S_1(\alpha_1, \alpha_2)=-\frac{\sinh[\frac{3}{2}(\alpha_1-\alpha_2)]-i}{\sinh[\frac{3}{2}(\alpha_1-\alpha_2)]+i}.\label{ai8}
\end{eqnarray}

(ii) Next, we consider the scattering of two second kind of elementary excitations. There are two 2-strings and
we denote them as $\lambda_{N-4}=\alpha_1-\frac{i\pi}{3}$, $\lambda_{N-3}=\alpha_1+\frac{i\pi}{3}$,
$\lambda_{N-2}=\alpha_2-\frac{i\pi}{3}$ and $\lambda_{N-1}=\alpha_2+\frac{i\pi}{3}$, where both $\alpha_1$ and $\alpha_2$ are real.
The rest zero points are real. Then the BAEs can be expressed as
\begin{eqnarray}
&&\left[\frac{\sinh(\lambda_j-\frac{i}{6}\pi)}
{\sinh(\lambda_j+\frac{i}{6}\pi)}\right]^N
= -\prod_{k=1}^{N-5}
\frac{\sinh(\lambda_j-\lambda_k+\frac{i}{3}\pi)}
{\sinh(\lambda_j-\lambda_k-\frac{i}{3}\pi)}
\prod_{l=1}^{2}
\frac{\sinh(\lambda_j-\alpha_l+\frac{2i}{3}\pi)}
{\sinh(\lambda_j-\alpha_l-\frac{2i}{3}\pi)},\nonumber \\
&& \qquad j=1,\cdots,N-5, \label{E22_10} \\
&&\left[\frac{\sinh(\alpha_m+\frac{i}{6}\pi)}
{\sinh(\alpha_m-\frac{i}{6}\pi)}\right]^N
= \prod_{k=1}^{N-5}
\frac{\sinh(\alpha_m-\lambda_k+\frac{2i}{3}\pi)}
{\sinh(\alpha_m-\lambda_k-\frac{2i}{3}\pi)}
\frac{\sinh(\alpha_m-\alpha_p+\frac{i}{3}\pi)}
{\sinh(\alpha_m-\alpha_p-\frac{i}{3}\pi)},\nonumber \\
&& \qquad m=1,2, \quad p=1,2, \quad m\neq p.\label{E22_1}
\end{eqnarray}
Taking the logarithm of Eq.\eqref{E22_10}, we have
\begin{eqnarray}\label{logE22}
\theta_1(\lambda_j) = \frac{2\pi I_j}{N}
-\frac{1}{N}\sum_{k=1}^{N-5}\theta_2(\lambda_j-\lambda_k)
+\frac{1}{N}\theta_2(\lambda_j-\alpha_1)+\frac{1}{N}\theta_2(\lambda_j-\alpha_2).
\end{eqnarray}
In the thermodynamic limit, the corresponding integral equation reads
\begin{eqnarray}
&&\hspace{-1cm}\rho_{22}(\lambda)+\rho_{22}^h(\lambda)=a_1(\lambda)+
\int_{-\infty}^{\infty}a_2(\lambda-\mu)\rho_{22}(\mu)d\mu
-\frac{1}{N}a_2(\lambda-\alpha_1)-\frac{1}{N}a_2(\lambda-\alpha_2), \label{s0} \\
&&\hspace{-1cm}\rho_{22}^h(\lambda)=\frac{1}{3N}\delta(\lambda-\lambda_0^h)
+\frac{1}{3N}\delta(\lambda+\lambda_0^h)+\frac{1}{N}\delta(\lambda-\lambda_1^h)
+\frac{1}{N}\delta(\lambda-\lambda_2^h).\label{s}
\end{eqnarray}
The density of the zero points $\rho_{22}(\lambda)$ satisfies
\begin{eqnarray}
\int_{-\infty}^{\infty}\rho_{22}(\lambda)d\lambda=\frac{N-5}{N}.
\end{eqnarray}
With the help of Fourier transform, the solution of Eq.\eqref{s0} is
\begin{eqnarray}\label{rho22w}
&&\tilde{\rho}_{22}(w)=
\frac{1}{1-\tilde{a}_2(w)}\Big\{
\tilde{a}_1(w)-\frac{e^{iw\lambda_0^h}}{3N}
-\frac{e^{iw\lambda_0^h}}{3N}+\frac{1}{N}e^{iw\alpha_1}\tilde{a}_4(w)
\nonumber\\
&&\qquad\qquad\quad
+\frac{1}{N}e^{iw\alpha_2}\tilde{a}_4(w)
-\frac{1}{N}e^{iw\lambda_1^h}
-\frac{1}{N}e^{iw\lambda_2^h}\Big\}\nonumber\\
&&\qquad\quad\; =\tilde{\rho}_g(w)+\delta\tilde{\rho}_2(w)_{\alpha_1,\lambda^h_1} +\delta\tilde{\rho}_2(w)_{\alpha_2,\lambda^h_2},
\end{eqnarray}
where $\delta\tilde{\rho}_2(w)_{\alpha,\lambda^h}$ is given by~\eqref{drho2w} with the replacing of $\lambda_2^h$ by $\lambda^h$.
Thus the density of zero points $\rho_{22}(\lambda)$ is
\begin{eqnarray}\label{rho22}
\rho_{22}(\lambda)=\rho_g(\lambda)+\delta\rho_2(\lambda)_{\alpha_1,\lambda^h_1} +\delta\rho_2(\lambda)_{\alpha_2,\lambda^h_2},
\end{eqnarray}
where $\delta\rho_2(\lambda)_{\alpha,\lambda^h}$ is given by~\eqref{drho2} with the replacing of $\lambda_2^h$ by $\lambda^h$.

In the thermodynamic limit, the BAEs \eqref{E22_1} can be rewritten as
\begin{eqnarray}
\hspace{-0.2cm}\left[\frac{\sinh(\frac{i}{6}\pi-\alpha_m)}
{\sinh(\frac{i}{6}\pi+\alpha_m)}\right]^N
=e^{Ni\int_{-\infty}^{\infty}\theta_4(\alpha_m-\lambda)\rho_{22}(\lambda)d\lambda}
\frac{\sinh[\frac{i}{3}\pi-(\alpha_m-\alpha_p)]}{\sinh[\frac{i}{3}\pi+(\alpha_m-\alpha_p)]}, m,p=1,2; m\neq p.\label{E22_2}
\end{eqnarray}
As we have mentioned, the position of hole $\lambda_m^h$ is equal to $\alpha_m$. Substituting Eq.\eqref{rho22} into~\eqref{E22_2} and taking the integral, we have
\begin{eqnarray}
\left[\frac{i-\sqrt{2}\sinh(\frac{3}{2}\alpha_m)}{i+\sqrt{2}\sinh(\frac{3}{2}\alpha_m)}\right]^N =\frac{i-\sinh[\frac{3}{2}(\alpha_m-\alpha_p)]}{i+\sinh[\frac{3}{2}(\alpha_m-\alpha_p)]}, \;\;m,p=1,2,\;\; m\neq p.
\end{eqnarray}
Then the two-body scattering matrix is
\begin{eqnarray}
S_2(\alpha_1, \alpha_2)=-\frac{\sinh[\frac{3}{2}(\alpha_1-\alpha_2)]-i}{\sinh[\frac{3}{2}(\alpha_1-\alpha_2)]+i},\label{ai9}
\end{eqnarray}
which is the same as that in the first kind of elementary excitations.

(iii) Last, we consider the scattering between the first and the second kinds of elementary excitations.
We set $\lambda_{N-3}=\alpha_1-\frac{i\pi}{2}$, $\lambda_{N-2}=\alpha_2-\frac{i\pi}{3}$ and $\lambda_{N-1}=\alpha_2+\frac{i\pi}{3}$,
where $\alpha_1$, $\alpha_2$ and the rest zero points are real. In this case, the BAEs read
\begin{eqnarray}
&&\left[\frac{\sinh(\frac{i}{6}\pi-\lambda_j)}
{\sinh(\frac{i}{6}\pi+\lambda_j)}\right]^N
= \prod_{k=1}^{N-4}
\frac{\sinh[\frac{i}{6}\pi+(\lambda_j-\lambda_k)]}
{\sinh[\frac{i}{6}\pi-(\lambda_j-\lambda_k)]}
\frac{\sinh[\frac{i}{6}\pi-(\lambda_j-\alpha_1)]}
{\sinh[\frac{i}{6}\pi+(\lambda_j-\alpha_1)]}\nonumber\\
&&\hspace{1cm}\times\frac{\sinh[\frac{2i}{3}\pi+(\lambda_j-\alpha_2)]}
{\sinh[\frac{2i}{3}\pi-(\lambda_j-\alpha_2)]},\quad j=1,\cdots, N-4,\label{E12_0} \\
&&\left[\frac{\sinh(\frac{i}{3}\pi+\alpha_1)}
{\sinh(\frac{i}{3}\pi-\alpha_1)}\right]^N
=\prod_{k=1}^{N-4}
\frac{\sinh[\frac{i}{6}\pi-(\alpha_1-\lambda_k)]}
{\sinh[\frac{i}{6}\pi+(\alpha_1-\lambda_k)]}
\frac{\sinh[\frac{i}{6}\pi+(\alpha_1-\alpha_2)]}
{\sinh[\frac{i}{6}\pi-(\alpha_1-\alpha_2)]},\label{E12_1}  \\
&&\left[\frac{\sinh(\frac{i}{6}\pi-\alpha_2)}
{\sinh(\frac{i}{6}\pi+\alpha_2)}\right]^N
=\prod_{k=1}^{N-4}
\frac{\sinh[\frac{2i}{3}\pi-(\alpha_2-\lambda_k)]}
{\sinh[\frac{2i}{3}\pi+(\alpha_2-\lambda_k)]}
\frac{\sinh[\frac{i}{6}\pi-(\alpha_2-\alpha_1)]}
{\sinh[\frac{i}{6}\pi+(\alpha_2-\alpha_1)]}.\label{E12_2}
\end{eqnarray}
Taking the logarithm of Eqs.\eqref{E12_0}, we have
\begin{eqnarray}\label{logE12}
\theta_1(\lambda_j) = \frac{2\pi I_j}{N}
-\frac{1}{N}\sum_{k=1}^{N-4}\theta_2(\lambda_j-\lambda_k)
+\frac{1}{N}\theta_1(\lambda_j-\alpha_1)+\frac{1}{N}\theta_2(\lambda_j-\alpha_2).
\end{eqnarray}
The corresponding integral equation in the thermodynamic limit is
\begin{eqnarray}
&&\hspace{-1cm}\rho_{12}(\lambda)+\rho_{12}^h(\lambda)=a_1(\lambda)+
\int_{-\infty}^{\infty}a_2(\lambda-\mu)\rho_{12}(\mu)d\mu
-\frac{1}{N}a_1(\lambda-\alpha_1)-\frac{1}{N}a_2(\lambda-\alpha_2),\label{0030}  \\
&&\hspace{-1cm}\rho_{12}^h(\lambda) = \frac{1}{3N}\delta(\lambda-\lambda_0^h)
+\frac{1}{3N}\delta(\lambda+\lambda_0^h)+\frac{1}{N}\delta(\lambda-\lambda_2^h).\label{003}
\end{eqnarray}
The density of zero points $\rho_{12}(\lambda)$ satisfies
\begin{eqnarray}
\int_{-\infty}^{\infty}\rho_{12}(\lambda)d\lambda=\frac{N-4}{N}.
\end{eqnarray}
With the help of Fourier transform, the solution of Eq.\eqref{0030} is
\begin{eqnarray}
&&\tilde{\rho}_{12}(w)=
\frac{1}{1-\tilde{a}_2(w)}\Big\{
\tilde{a}_1(w)-\frac{e^{iw\lambda_0^h}}{3N}
-\frac{e^{iw\lambda_0^h}}{3N}-\frac{e^{iw\alpha_1}\tilde{a}_1(w)}{N}
-\frac{e^{iw\alpha_2}\tilde{a}_2(w)}{N}
-\frac{e^{iw\lambda_2^h}}{N}\Big\}\nonumber\\[6pt]
&&\qquad\quad\equiv \tilde{\rho}_g(w)+\delta\tilde{\rho}_1(w)_{\alpha_1} +\delta\tilde{\rho}_2(w)_{\alpha_2,\lambda^h_2}.
\end{eqnarray}
Then the density of zero points $\rho_{12}(\lambda)$  is given by
\begin{eqnarray}\label{rho12}
\rho_{12}(\lambda)=\rho_g(\lambda)+\delta\rho_1(\lambda)_{\alpha_1} +\delta\rho_2(\lambda)_{\alpha_2,\lambda^h_2}.
\end{eqnarray}
In the thermodynamic limit, the BAE \eqref{E12_1} reads
\begin{eqnarray}
\left[\frac{\sinh(\frac{i}{3}\pi+\alpha_1)}
{\sinh(\frac{i}{3}\pi-\alpha_1)}\right]^N
=e^{Ni\int_{-\infty}^{\infty}\theta_1(\alpha_1-\lambda)\rho_{12}(\lambda)d\lambda}
\frac{\sinh[\frac{i}{6}\pi+(\alpha_1-\alpha_2)]}{\sinh[\frac{i}{6}\pi-(\alpha_1-\alpha_2)]}. \label{E12_11}
\end{eqnarray}
Substituting Eq.\eqref{rho12} into~\eqref{E12_11} and taking the integral, we have
\begin{eqnarray}
\left[\frac{i-\sinh(\frac{3}{2}\alpha_1)}{i+\sinh(\frac{3}{2}\alpha_1)}\right]^N =\left[\frac{i-\sqrt{2}\sinh[\frac{3}{2}(\alpha_1-\alpha_2)]} {i+\sqrt{2}\sinh[\frac{3}{2}(\alpha_1-\alpha_2)]}\right]^2.
\end{eqnarray}
In the thermodynamic limit, we rewrite the BAE \eqref{E12_2} as
\begin{eqnarray}
\left[\frac{\sinh(\frac{i}{6}\pi-\alpha_2)}
{\sinh(\frac{i}{6}\pi+\alpha_2)}\right]^N
=e^{Ni\int_{-\infty}^{\infty}\theta_4(\alpha_2-\lambda)\rho_{12}(\lambda)d\lambda}
\frac{\sinh[\frac{i}{6}\pi-(\alpha_2-\alpha_1)]}{\sinh[\frac{i}{6}\pi+(\alpha_2-\alpha_1)]} .\label{E12_22}
\end{eqnarray}
Substituting Eq.\eqref{rho12} into~\eqref{E12_22} and taking the integral, we obtain
\begin{eqnarray}
\left[\frac{i-\sqrt{2}\sinh(\frac{3}{2}\alpha_2)}{i+\sqrt{2}\sinh(\frac{3}{2}\alpha_2)}\right]^N =\frac{i-\sqrt{2}\sinh[\frac{3}{2}(\alpha_2-\alpha_1)]} {i+\sqrt{2}\sinh[\frac{3}{2}(\alpha_2-\alpha_1)]}.
\end{eqnarray}
Then the two-body scattering matrix is
\begin{eqnarray}
S_3(\alpha_2, \alpha_1)=-\frac{\sqrt{2}\sinh[\frac{3}{2}(\alpha_2-\alpha_1)]-i} {\sqrt{2}\sinh[\frac{3}{2}(\alpha_2-\alpha_1)]+i}=S^{-1}_3(\alpha_1, \alpha_2).\label{ai10}
\end{eqnarray}
The scattering matrix \eqref{ai10} is different from the previous results \eqref{ai8} and \eqref{ai9}.

\section{Conclusions}
\label{sec:concluding remarks}

In this paper, we have studied the thermodynamic limit of the antiperiodic XXZ spin chain with the anisotropic parameter $\eta=\frac{\pi i}{3}$, in which  particular case roots of the transfer matrix satisfy the homogeneous BAEs (\ref{BAE-2}). We can parameterize eigenvalues of the transfer matrix by their zero points instead of Bethe roots. We obtain patterns of distribution of their zero points. Based on them, we calculate the ground state energy and check the analytical results numerically.
We find that the density of ground state energy is equal to that with the periodic boundary condition.
The system has two types of elementary excitations and we compute the corresponding excited energies in the thermodynamic limit.
If there are more than one elementary excitations, we obtain the two-body scattering matrix of quasi-particles.
We find that there are three types of scattering processes. Then we discussed them separately.

\section*{Acknowledgments}

We would like to thank Professor Y. Wang for his valuable discussions and continuous encouragement.
The financial supports from the National Natural Science Foundation of China (Grant Nos. 12074410, 12147160, 12047502, 11934015, 11975183, 11947301 and 11774397), Major Basic Research Program of Natural Science of Shaanxi Province (Grant Nos. 2021JCW-19 and 2017ZDJC-32), Australian
Research Council (Grant No. DP 190101529), the Strategic
Priority Research Program of the Chinese Academy of Sciences (Grant No. XDB33000000),
the fellowship of China Postdoctoral Science Foundation (Grant No. 2020M680724), and the
Double First-Class University Construction Project of
Northwest University are gratefully acknowledged.

\end{document}